\documentstyle [preprint,prl,aps] {revtex}
\begin {document}

\title {Excited-state relaxations and Franck-Condon shift \break \hfill
in Si quantum dots}
\author {A. Franceschetti and S.T. Pantelides}
\address {Department of Physics, Vanderbilt University, Nashville, TN, and 
Solid State Division, Oak Ridge National Laboratory, Oak Ridge, TN} 
\maketitle

\begin {abstract}
Excited-state relaxations in molecules are responsible for a red shift of the absorption
peak with respect to the emission peak (Franck-Condon shift).
The magnitude of this shift in semiconductor quantum dots is still unknown.
Here we report first-principle calculations of excited-state relaxations
in small (diameter $ \le 2.2 \; {\rm nm}$) Si nanocrystals, 
showing that the Franck-Condon shift is surprisingly large ($\sim 60 \; {\rm meV}$
for a $2.2 \; {\rm nm}$-diameter nanocrystal).
The physical mechanism of the excited-state relaxations changes abruptly 
around $\sim 1 \,$ nanometer in size, 
providing a clear demarcation between ``molecules'' and ``nanocrystals''.
\end {abstract}

\newpage 

The Stokes shift commonly observed in molecules \cite {Herzberg} and ionic solids \cite {Rebane},
has its origin in excited-state atomic relaxations.
When an electron-hole pair is created by optical excitation, the final state has approximately 
the same atomic configuration as the initial state (Franck-Condon principle). 
Prior to emission, however, the system can relax to a new configuration with lower total energy.
Recombination occurs from the relaxed atomic configuration, 
leading to a red-shift of the emission lines with respect to the absorption lines
(Franck-Condon shift).  The Franck-Condon shift in molecules and ionic
solids can be as large as several eV.
The magnitude of excited-state relaxations
in semiconductor quantum dots, on the other hand, is still controversial.
Quantum dots grown by colloidal chemistry methods
range in size between between $10^2$ and $10^4$ atoms, 
so excited-state relaxations could be significant. 
Continuum models based on the effective-mass approximation, however,
have predicted a Franck-Condon shift of only a few meV \cite {Martin,Takagahara}.

A schematic diagram of the relevant electronic energy levels as a function of the generic configuration
variable ${\bf R}$ is shown in Fig. 1. 
The minimum-energy atomic configuration of the quantum dot 
in the electronic ground state (GS) is different from the minimum-energy atomic configuration 
of the dot in the singlet (S) or triplet (T) excited states.
At low temperature and in the absence of light, the quantum dot is in the ground-state
geometry ${\bf R}_{GS}$.
The lowest-energy allowed optical transition takes the system into the singlet excited state,
as the lower-energy triplet state is optically inactive.
Since the excitation of an electron-hole pair occurs on a much faster time scale compared
to atomic vibrations, the atomic configuration immediately after
absorption is still the ground state configuration ${\bf R}_{GS}$, as indicated by the vertical
line in Fig. 1. Absorption can actually occur into several vibrational states associated
with the singlet electronic state, leading to a broadening of the absorption line.
The exciton relaxation then proceeds according to a few characteristic times:
(i) The spin-flip time $\tau_{flip}$, which is the time required for the exciton to switch
from the singlet state to the triplet state,
(ii) the recombination times $\tau_{rec}^{S,T}$ 
(with $\tau_{rec}^S \ll \tau_{rec}^{T}$), which include radiative and non-radiative recombination paths,
and (iii) the relaxation times $\tau_{rlx}^{S,T}$ in the singlet and triplet states, which are
the characteristic times for the dissipation of the vibrational energy.
The relaxation times $\tau_{rlx}$ are ultimately determined  by the coupling
of the quantum dot with the environment.
If $\tau_{rlx}^{S,T} \ll \tau_{rec}^T$ 
(and $\tau_{flip}  < \tau_{rec}^{T}$), the quantum dot relaxes to the lowest-energy
triplet excited-state configuration (${\bf R}_{XS}$) before the electron and the hole can recombine.
In this case the total Stokes shift is $\Delta E = \Delta E_{FC} + \Delta E_{ST}$,
where $\Delta E_{FC}$ is the Franck-Condon shift and $\Delta E_{ST}$ is the exciton exchange splitting.
Different vibrational states may be involved in 
the emission process, leading to a broadening of the emission peak \cite {Herzberg,Rebane}.

In this Letter we investigate, using {\it ab-initio} density-functional methods, 
the excited-state dynamics of Si quantum dots, and calculate the ensuing Franck-Condon shift.
We find that for small Si quantum dots ($1.0 - 2.2 \; {\rm nm}$ diameter)
the Franck-Condon shift is surprisingly large.
For example, in the case of a $ \sim 2.2 \; {\rm nm}$-diameter quantum dot 
we predict a Franck-Condon shift of $\sim 60 \; {\rm meV}$, 
versus an electron-hole exchange splitting of only $\sim 8 \; {\rm meV}$.
By analyzing the physical mechanism responsible for the Franck-Condon shift 
we are able to identify two physically distinct regimes: 
for sub-nanometer Si clusters the dominant
mechanism is the stretching of a single Si-Si bond upon electronic excitation, 
while for larger Si nanocrystals the Franck-Condon shift originates from a change in 
the overall shape of the nanocrystal in the presence of an electron-hole pair. 
This distinction provides a clear demarcation line between the molecular regime
and the nanocrystal regime.

The calculations were performed using {\it ab-initio} density-functional theory in the
local spin density (LSD) approximation. 
We used ultra-soft pseudopotentials to describe the electron-ion interaction,
and the plane-wave representation (with an energy cutoff of 150 eV)
to describe the Kohn-Sham orbitals.
Triplet excited states can be calculated within density-functional theory
by minimizing the total energy of the system in the triplet spin configuration.
We find that in practice this approach works very well. For example in the case of the silane molecule
SiH$_4$, we calculate a triplet excitation energy of $8.1 \; {\rm eV}$, compared to the value 
of $8.7 \; {\rm eV}$ obtained using diffusion Quantum Monte Carlo \cite {Grossman}.
In the bulk limit, we expect our approximation for the excited-state energy
to converge to the LSD band gap of Si, which is $\sim 0.68 \; {\rm eV}$ 
lower than the experimental band gap. So we expect the calculated excitation energies
to differ from the experimental excitation energies approximately by a constant energy shift
\cite {Delerue}.

Our approach to the calculation of excited states is applicable only to the lowest-energy triplet state
(see Fig. 1), where the electron excited to the LUMO and the electron remaining in 
the HOMO have parallel spin, i.e. $|T \rangle = |\uparrow \, \uparrow \rangle$.
Calculations of the excited-state singlet energy surface would require the handling
of a two-determinantal wave function, which is beyond the reach of simple density-functional theories
(see however Ref. \onlinecite{Frank}).
For comparison, we have also performed excited-state calculations on the ``mixed'' energy surface
given by $|M \rangle = |\uparrow \, \downarrow \rangle$. It can be shown that, for a given atomic
configuration, the energy difference $E^M - E^T$ is one half of the singlet-triplet splitting
$E^S - E^T$ (see Ref. \onlinecite{Frank}).

We consider here nearly spherical Si nanocrystals centered on a Si atom.
The initial atomic configuration (before atomic relaxations) is obtained by cutting out a sphere
from a bulk Si crystal. The Si-Si bond length is taken as the bulk LSD
bond length ($2.33 \; {\rm \AA}$).
The surface atoms having three dangling bonds are removed,
while the remaining surface dangling bonds are passivated by H atoms. 
The nanocrystals considered here range in size from 
29 Si atoms ($10.3 \; {\rm \AA}$ diameter) to 275 Si atoms ($21.7 \; {\rm \AA}$ diameter).

The calculation of the Franck-Condon shift is carried out in 4 steps:
(i) First, the ground-state atomic configuration is obtained by minimizing the total energy
of the nanocrystal with respect to the atomic positions, 
as dictated by quantum-mechanical forces. 
This step gives the ground-state total energy $E^{GS} ({\bf R}_{GS})$.
(ii) Then we excite an electron-hole pair in the triplet state, and calculate the excited-state energy
in the ground-state geometry: $E^{T} ({\bf R}_{GS})$.
The difference $E^{T} ({\bf R}_{GS}) - E^{GS} ({\bf R}_{GS})$ is the triplet excitation
energy $E_{exc}$ (not to be confused with the absorption energy $E_{abs}$
shown in Fig. 1).
(iii) Next, we relax the atomic positions on the triplet excited-state energy surface,
thus obtaining the excited-state total energy in the excited-state atomic configuration (XS):
$E^{T} ({\bf R}_{XS})$.
(iv) Finally, we calculate the ground-state total energy in the excited-state geometry:
$E^{GS} ({\bf R}_{XS})$. The energy difference 
$E^{T} ({\bf R}_{XS}) - E^{GS} ({\bf R}_{XS})$
is the emission energy $E_{emi}$.
The Franck-Condon shift $\Delta E_{FC}$ is then given by 

\begin {equation}
\Delta E_{FC} = \left [ E^{T} ({\bf R}_{GS}) - E^{GS} ({\bf R}_{GS}) \right ] - 
\left [ E^{T} ({\bf R}_{XS}) - E^{GS} ({\bf R}_{XS}) \right ] =
E_{exc} - E_{emi}.
\end {equation}

\noindent $\Delta E_{FC}$ can be further decomposed into an excited-state contribution 
$\Delta E_{XS} = E^{T} ({\bf R}_{GS}) - E^{T} ({\bf R}_{XS})$
and a ground-state contribution
$\Delta E_{GS} = E^{GS} ({\bf R}_{XS}) - E^{GS} ({\bf R}_{XS})$ (see Fig. 1).
Note that $\Delta E_{XS}$ and $\Delta E_{GS}$ are total-energy differences between
different atomic configurations of the same system. Thus, we expect
the accuracy of the calculated Franck-Condon shift to be comparable with the accuracy of
calculated  vibrational energies (a few \% error in the case of bulk Si).

The results for Si nanocrystals are summarized in Table I.
Note that, in the size range considered here, the ground-state contribution to the Franck-Condon shift
$\Delta E_{GS}$ is larger than the excited-state 
contribution $\Delta E_{XS}$, particularly for the smaller nanocrystals. 
This difference reflects the reduced curvature and increased non-parabolicity
of the excited state energy surface compared to the ground-state energy surface. 
Martin {\it et al.} \cite {Martin} proposed a simple model, 
based on the envelope-function approximation and empirical deformation potentials,
to estimate the Franck-Condon shift in Si nanocrystal. 
They predicted that for a nanocrystal with an excitonic gap of $2.3 \; {\rm eV}$
(corresponding to the largest nanocrystal considered here, see Table I)
the Franck-Condon shift would range from 9 to 21 meV, depending on the parameters of the model.
Using a similar continuum model, Takagahara {\it et al.} \cite {Takagahara}
obtained a Stokes shift of $\sim 7 \; {\rm meV}$ for the same nanocrystal.
Our {\it ab-initio} calculations show that these models significantly underestimate the 
Franck-Condon shift of Si quantum dots.

The Franck-Condon shift is very large
in the small Si$_{29}$ H$_{36}$ cluster, 
where the distortions due to the electronic
excitation are large. 
Figure 2(a) shows the Si-Si bond-length distribution $n(L)$
of this cluster, both in the ground-state geometry
and in the relaxed excited-state (triplet) geometry. 
We see that $n(L)$ is similar in the
ground state and in the excited state, 
except for a single Si-Si bond that is stretched 
by about $15 \%$ in the excited-state geometry.
The HOMO and LUMO wave functions in the excited-state geometry
are strongly localized around the stretched Si-Si bond, 
as shown in the insets in Fig. 2(a), and the corresponding
energy levels are well inside the band gap of the Si nanocrystal,
thus determining the large Franck-Condon shift of $2.9 \; {\rm eV}$.
Using tight-binding total energy calculations
Allan {\it et al.} \cite {Allan} predicted that in small hydrogen-passivated 
Si nanocrystals excitons can become self-trapped in a (meta)stable state localized at the surface. 
Our ab-initio calculations show that excited-state relaxations lead to the 
spontaneous formation of a stretched bond in the interior of the nanocrystal,
and therefore this effect should depend weakly on the type of surface passivation.
Hirao \cite {Hirao} calculated the Stokes shift
of Si$_{29}$ H$_{36}$ nanocrystals, finding a value of $0.22 \; {\rm eV}$,
over an order of magnitude smaller than our result. This difference may be due to the lower
energy cutoff used in the plane-wave expansion of Ref. \onlinecite {Hirao}.

Figure 2(b) shows the bond-length distribution of the Si$_{87}$ H$_{76}$ nanocrystal.
The distribution is centered around the bulk Si-Si bond length, with very little
differences between the ground state and the excited state.
In fact, we find that the Franck-Condon shift in this nanocrystal (and in larger nanocrystals)
is due to a change in the overall shape of the nanocrystal, from spherical to ellipsoidal, 
upon electronic excitation.
This change of shape leads to a splitting of the states at the top of the valence band (which
are degenerate in the T$_d$ representation) and thus to the Franck Condon shift.
The insets in Fig. 2(b) show that 
the electron and hole wave functions
in the excited-state configuration are delocalized over the entire nanocrystal.

As shown in Fig. 1, the exchange contribution to the Stokes shift is given
(at low temperatures) by the singlet-triplet splitting $\Delta E_{ST}$ evaluated at
the ground-state geometry $\{ {\bf R}_{GS} \}$.
We therefore calculate the exchange contribution as:

\begin {equation}
\Delta E_{ST} =  2 \left [ E^{M} ({\bf R}_{GS}) - E^{T} ({\bf R}_{GS}) \right ]. 
\end {equation}

\noindent The results are shown in the last column of Table I. We find that
the electron-hole exchange splitting is significantly smaller than the Franck-Condon shift,
even for the largest nanocrystal considered here. 
The singlet-triplet splitting of Si quantum dots
had been calculated in the past using a variety of empirical approaches, 
including the effective-mass approximation \cite {Calcott},
the tight-binding approximation \cite {Martin}, 
and the semi-empirical pseudopotential method \cite {Reboredo}. 
Our {\it ab-initio}  calculations are in good agreement with 
the pseudopotential calculations of Ref. \onlinecite {Reboredo}.

The Stokes shift of Si nanocrystals has been measured using both optical and thermal
methods \cite {Calcott,Kovalev,Brongersma}. 
It is difficult to compare our results directly with experimental data,
because Si nanocrystals synthesized by colloidal chemistry techniques are typically
larger than those considered here, and because the absorption coefficient at the absorption edge
is small. 
Using selective laser excitation Calcott {\it et al.} \cite {Calcott}
and Kovalev {\it et al.} \cite {Kovalev}
measured the red shift of the photoluminescence onset with respect to the excitation energy.
For example, for an excitation energy of $2.41 \; {\rm eV}$, 
Calcott {\it et al.} \cite {Calcott} measured a red shift of $23 \; {\rm meV}$.
The photoluminescence onset was attributed to zero-phonon emission,
and the observed Stokes shift was interpreted as due to the electron-hole exchange splitting.
Our calculations suggest that in small Si nanocrystals the Franck-Condon shift 
becomes progressively more important, until it dominates over the exchange splitting.

In conclusion, we have shown by excited-state density-functional calculations
that the Franck-Condon shift in small Si nanocrystals
is larger than previously thought.
We have found that in sub-nanometer clusters the Franck-Condon shift originates
from the stretching of a Si-Si bond, while in larger nanocrystals it is due
to a change in the overall shape of the nanocrystal upon electron-hole excitation.

This work was supported in part by
the DOE Computational Materials Science Network grant DE-FG02-02ER45972,
NSF grant DMR9803768, 
the US DOE under contract DE-AC05-00OR22725 with the
Oak Ridge National Laboratory, managed by UT-Battelle, LLC, and the
William A. and Nancy F. McMinn Endowment at Vanderbilt University.

\begin {table}
\caption {Calculated Franck-Condon shift $\Delta E_{FC}$ 
and electron-hole exchange splitting $\Delta E_{ST}$ (in eV)
of a few hydrogen-passivated Si nanocrystals. Also shown are the ground-state and excited-state
contributions to the Franck-Condon shift: $\Delta E_{FC} = \Delta E_{XS} + \Delta E_{GS}$.}

\begin {tabular} {lccccccc}
Nanocrystal & Diameter (${\rm \AA}$) & 
$\Delta E_{XS}$ & $\Delta E_{GS}$ & $\Delta E_{FC}$ & $\Delta E_{ST}$ \\
\tableline
Si$_{29}$ H$_{36}$   & 10.3 & 0.79 & 2.13 & 2.92 & 0.051 \\
Si$_{87}$ H$_{76}$   & 14.8 & 0.12 & 0.20 & 0.32 & 0.021 \\
Si$_{147}$ H$_{100}$ & 17.6 & 0.09 & 0.12 & 0.21 & 0.014 \\
Si$_{275}$ H$_{172}$ & 21.7 & 0.03 & 0.03 & 0.06 & 0.008 \\
\end {tabular}
\end {table}

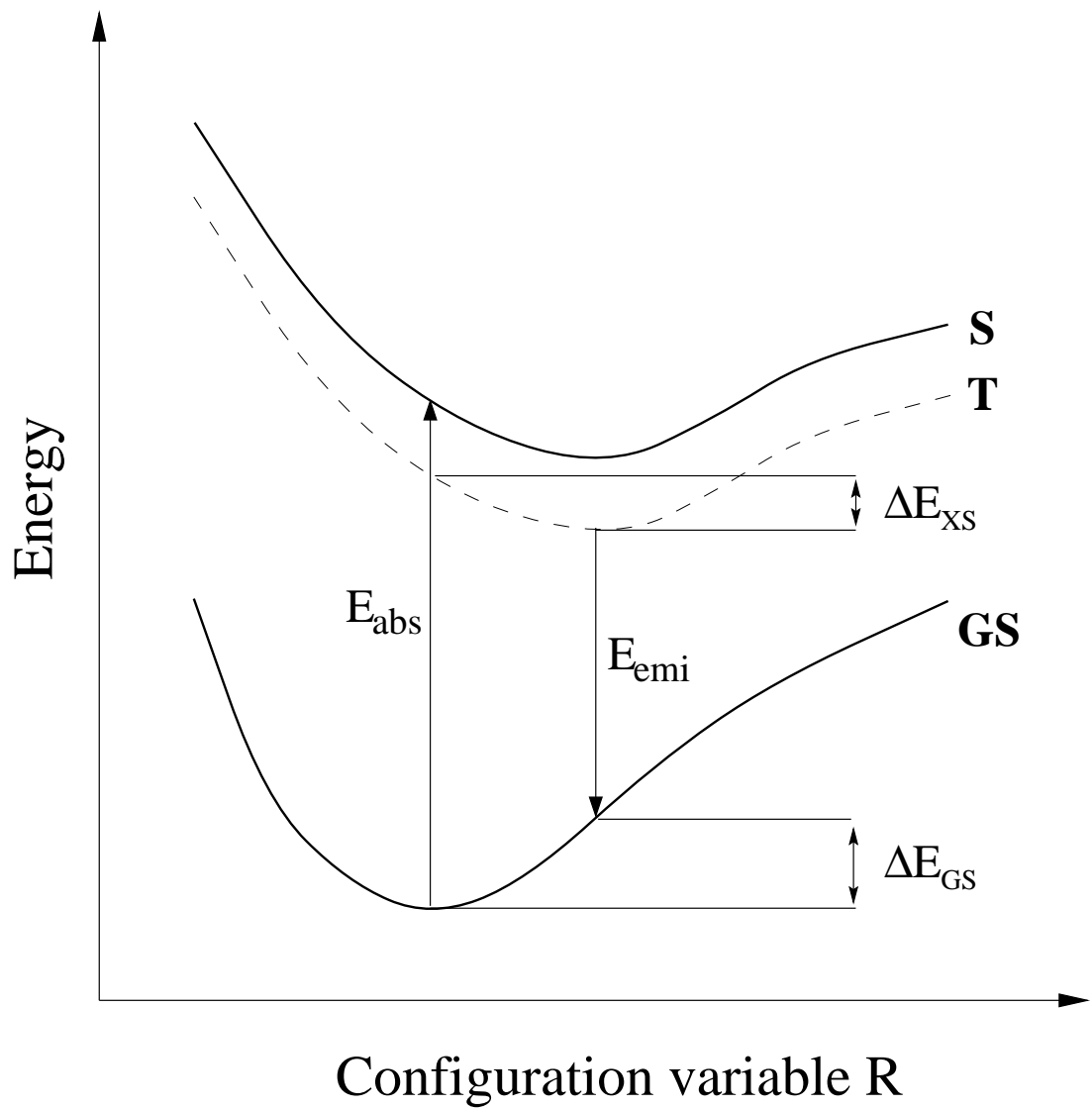
\begin {figure}
\caption {Schematic diagram of the ground-state (GS) and excited-state singlet (S) and triplet (T) 
energy surfaces of a semiconductor quantum dot. Light is absorbed ($E_{abs}$) by exciting 
the quantum dot from the ground state to the optically active singlet state.
Emission occurs ($E_{emi}$) from the optically inactive triplet state, leading
to a resonant Stokes shift of the emission line. The Franck-Condon contribution to the Stokes
shift is the sum of $\Delta E_{XS}$ and $\Delta E_{GS}$.}
\end {figure}

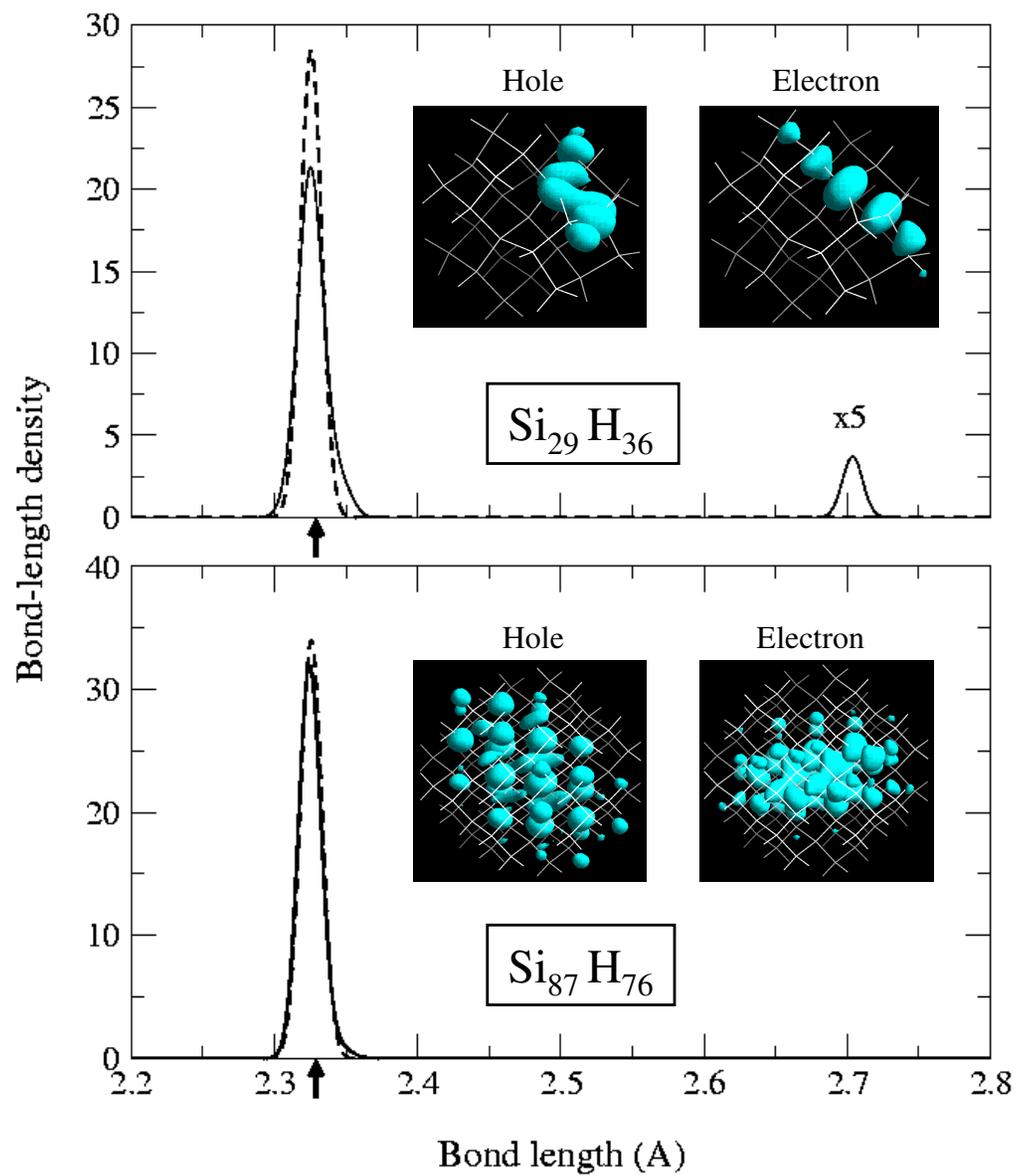
\begin {figure}
\caption {Bond-length distribution in the ground-state geometry (dashed lines)
and in the triplet excited-state geometry (solid lines) for the Si$_{29}$H$_{36}$ nanocrystal
and the Si$_{87}$H$_{76}$ nanocrystal. The insets show the electron (LUMO) and hole (HOMO)
single-particle wave functions in the excited-state configuration. The arrows denote
the calculated bulk Si-Si bondlength.}
\end {figure}

\end {document}